\documentclass[a4paper,prd,twocolumn,nofootinbib,superscriptaddress,floatfix]{revtex4}
\usepackage{cancel}
\usepackage{graphicx}
\usepackage{epsfig}
\usepackage{amsmath,amsfonts,amssymb}
\usepackage{float}
\usepackage{subfig}
\usepackage{mathtools}          

\begin{document}


\title{Can Complex Collective Behaviour Be Generated Through Randomness, Memory and a Pinch of Luck?}

\author{Pedro M. F. Pereira}

\affiliation{Departamento de F\'{\i}sica, Instituto Superior T\'ecnico, Universidade de Lisboa, Lisboa, Portugal\\
pedromanuelpereira@tecnico.ulisboa.pt}


\begin{abstract}
Machine Learning techniques have been used to teach computer programs how to play games as complicated as Chess and Go. These were achieved using powerful tools such as Neural Networks and Parallel Computing on Supercomputers. In this paper, we define a model of populational growth and evolution based on the idea of Reinforcement Learning, but using only the 3 sources stated in the title processed on a low-tier laptop.\\
The model correctly predicts the development of a population around food sources and their migration in search of a new one when the known ones become saturated. Additionally, we compared our model to a pure random one and the population number was fitted to a logistic function for two interesting evolutions of the system.
\end{abstract}

\maketitle

\section{Introduction}

The use of previous experiences to influence future actions is known in the literature as Reinforcement Learning with Experience Replay \cite{reinforcementlearning} and can be schematized as follows:\\
\begin{figure}[H]
\centering
  \includegraphics[scale=0.3]{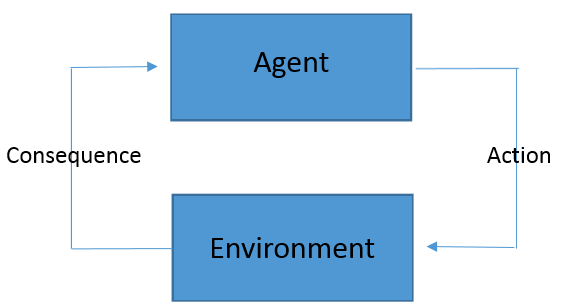}
\caption{Canonical Case of Reinforcement Learning}
  \label{fig:1}
\end{figure}

This area's main goal is to emulate the way of thinking of an intelligent being. It has been severely tested in artificial intelligence and game theory  and is behind AlphaGo, the AI that beat the world champion of the board game Go. \cite{go}\\
The aim of this work is to use this idea in a minimalist way, using the minimum of tools possible (without Markov Chains, Neural networks or parallel computing), guided only by logic and naturality.
Furthermore, the model used is different from the above, since the agent who makes the action is not the one receiving the information about the consequences (as it happens when the agent is learning how to play a game).\\

The premise behind this work is that it is possible to achieve complex collective behaviour with an initial population ($N$) where every member (named a \textit{cell} from now on) emulates a life form, thus needing to eat and being able to reproduce. At t=0, every \textit{cell} is a random walker - as in \cite{randwalk}. As time goes by the individual behaviour starts to deviate from this due to the creation and access to what I called \textit{memory}.  As it will become clear in the next pages,  a \textit{cell} will only have access to information created by its ancestors (with the exception of the initial population). 

A system like this is highly chaotic and thus a certain tuning of the initial conditions and set up variables is needed to generate something interesting. This is what I meant with a \textit{pinch of luck} in the title, as If I just set every parameter to random, an interesting evolution would be possible, but highly unlikely. Furthermore, it is also necessary to have parameters in a certain range, otherwise we get non physical  situations (like cells who can live almost all their life without eating or reproducing).\\
The simulations were done in a 2D square Lattice of side L and every iteration is named a day. The relevant set up variables are:\\

\begin{table}[H]
\begin{center}
\resizebox{\columnwidth}{!}{%
\begin{tabular}{ |c|c|c|}
 \hline
 \hline
 \textbf{Name} & \textbf{Label} & \textbf{Meaning} \\
 \hline
 \hline
Maximum Age & \textit{maxage} & Number of days a cell can live before dying of "old age"\\ 
 \hline
Maximum Days Without Food & \textit{maxswof} & Number of days a cell can live without reaching a food site.\\ 
 \hline
Minimun Age for Reproduction &\textit{minagerep}  & Minimum age for a cell to start reproducing.\\
 \hline
Daily Food Limit & \textit{dailyflimit} & Maximum number of cells who can feed on a food site daily.\\

   \hline \hline
\end{tabular}
}
\end{center}
\caption{Set Up Variables}
\label{fitcaractvi}
\end{table}

The initial conditions are the number of \textit{cells} at t=0, their spatial distribution and the spatial distribution and ratio of food and death sites. Food sites (green squares in the lattice) are places that when reached by a \textit{cell} set their \textit{hunger} variable to zero. Everyday that a \textit{cell} has not eaten adds one to their hunger variable, and when their hunger variable equals maxswof, the \textit{cell} dies.\\
Death sites (red squares in the lattice) are places that when reached by a \textit{cell} lead to their immediate death. They pretend to simulate an encounter with a predator or a zone with organic matter that is poisonous to the population.

Like any population growth model (see \cite{logistic}, for instance),  - since there will be a limit where every food site will overflow the population number will have to adjust to a logistic function (\ref{eq:1}):\\

\begin{equation}
N(d)=N_i + \frac{N_{eq}-N_i}{1+e^{-k(d-d_0)}}
\label{eq:1}
\end{equation}
where  $N_i$ is the initial number of \textit{cells} , $N_{eq}$ is the equilibrium number of \textit{cells}, $k$ is the steepness of the curve and $d_0$ is x-value of the function's midpoint.\\

\section{Population Behaviour Model}
The model consists in the implementation of the memory and the choices made by me when implementing the reproduction and movement of \textit{cells}. First, the basics (without memory) and the reproduction algorithm.\\

Every \textit{cell} starts in a square in the lattice and at the beginning of each day can move to any adjacent square, or stay in the same square as it is. The choice of where to move is random, meaning we have a spatial Equidistribution of Probability - random walker.\\
\begin{figure}[H]
\centering
  \includegraphics[scale=0.3]{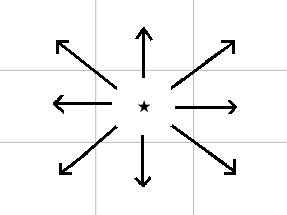}
\caption{Available Moves}
  \label{fig:2}
\end{figure}

\textit{Cells} have gender (0 or 1) and if two \textit{cells} of opposite genders are in the same square in the end of one day, and are older than \textit{minagerep}, they can reproduce. However, only one reproduction per \textit{cell} per day is allowed. Not fixing this to a reasonable number would lead to non physical growths.\\
Also, it is important to define a generation. \textit{Cells} from the initial population are generation 0 and the generation of a \textit{cell} $C$, offspring of \textit{cell} $A$ and $B$ is defined as:
\begin{align*}
gen_C=\max(gen_A,gen_B)+1
\end{align*}
Moreover, if one \textit{cell} has many options for reproduction it will choose at random one of the highest generation possible. This emulates the choice of the most adapted partner as it exists in nature, since a member of the highest generation will have a more developed memory.\\
Furthermore, if many \textit{cells} are in the same food site, the ones from higher generations and with lower age have priority to eat. This replicates what also happens in nature, with parents prioritizing their offspring life over theirs.

\begin{figure}[H] 
\centering
  \fbox{
    \centering
    \includegraphics[width=0.7\linewidth]{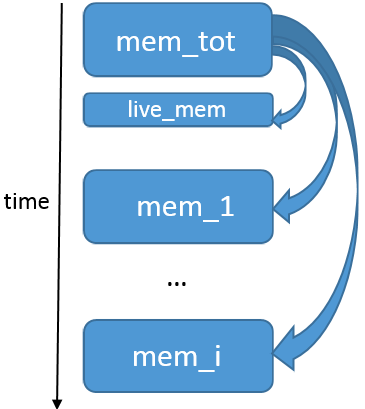}
    }

    \label{fig:3a} 
    \vspace{4ex}

  \fbox{
    \centering
    \includegraphics[width=0.7\linewidth]{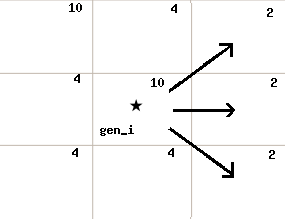} 
    }
  
    \label{fig:3b} 
    \vspace{4ex}
 
  \caption{(Top) Time evolution of the memories; (Bottom) The numbers on the top right corner of each square are the weights for generation i. By analyzing its surroundings, the \textit{cell} will choose one of the three moves with least weight associated.}
  \label{fig:3} 
\end{figure}

With memory, things get more complex. It was fundamental to register not only bad outcomes. It was decided that this would be done recurring to position weights. For every death a position gets the weight 6. Every time a \textit{cell} eats in a position, it gets the weight 0, the adjacent squares to it get the weight 1 and the remaining lattice positions get the weight 2. This privileges the food site above all the other positions, and also discriminates between the positions nearby and faraway from it. The probability of a \textit{cell} moving to a position will be:

\[
  P_{moving}(s)=
  \begin{cases}
     P_i & , \ w_{tot}=0 \\
     P_i [ e^{(P_i-\frac{w(s)}{w_{tot}}) } ] & ,\ w_{tot}>0
  \end{cases}
\]

where, n is the number of available positions, $P_i$ is the inverse of n, $w(s)$ the total weight of position s, and $w_{tot}$ the sum over every s of $w(s)$.
   This probability distribution is inspired in the exponential distribution, and it is easy to check that it also is a distribution. The particularity of this distribution is that if a certain $w(s)$ contributes poorly to $w_{tot}$,  $P_{moving}(s)$ will be bigger than $P_i$. This is a property of the best available moves, and its the core of the moving algorithm. The \textit{cell} will choose one of the available options at random (s) and test if $P_{moving}(s)$ is bigger or equal than $P_i$. If it is, it moves there, if it's not it will think about another one, and so on, until it moves to one of the best options. See FIG. \ref{fig:3} for clarification. This is a type of quality function (or Q-function) as described in \cite{qfunction}.

The process of memory creation is as follows:

1.  A total memory  \textit{$mem_{tot}$} registers the position and sums the corresponding weights, depending if it is a death or eating event.\\
2. The first newborn of generation 1 gets a copy of the current state of \textit{$mem_{tot}$} to his own memory \textit{$mem_1$}. This defines the memory of generation 1 for once and for all.\\
3. Repeat for every generation from now on.

As generation 0 doesn't have an ancestor to create memories for them, it starts with a live-memory that is updated after the end of each day. This emulates the communication between the initial members of the population, and lasts only their lifetime.

One last and important detail of the model remains to explain. What happens when $dailyflimit$ is reached? There is still a margin due to $maxswof$, so the limit when \textit{cells} start dying near the food sites will actually be around the product of these two quantities. An intelligent life form would try to find another food source when it notices that the one it has been using is not enough anymore. Thus, when the weight of the food site equals the weight of a nearby sites (something impossible before \textit{cells} start dying of hunger) there is an inversion in the weights and from that time on, every time a \textit{cell} eats there it gets the weight 0, the adjacent squares to it get the weight 2 and the remaining lattice positions get the weight 1. This still privileges the food site above all the other positions, however it gives the information that the nearby places are not good anymore. \\
The weights system reverts to its initial state when the population number gets below the critical number (the number of \textit{cells} that exist when  the weight of the food site equals the weight of a nearby site).

 \section{Results and Discussion}
 The setup variables were set to:\\
1.$ maxage=20$\\
2. $maxswof=5$\\
3. $minagerep=5$\\
4. $dailyflimit=10$   \\
The initial conditions used for all the results were:\\

 \begin{figure}[H]
\centering
  \includegraphics[scale=0.4]{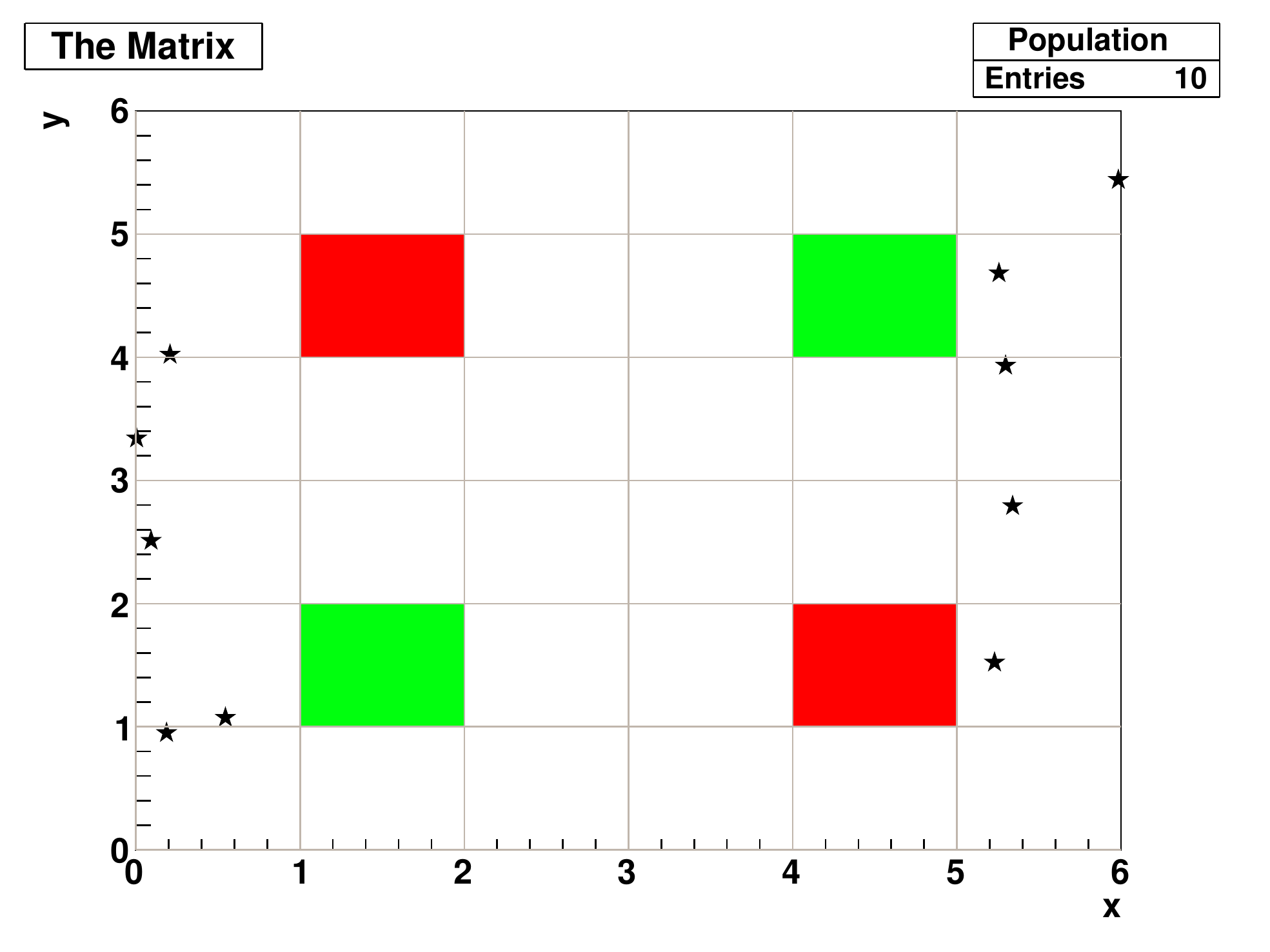}
\caption{Day 0 - Map with Fixed Food and Death Sites and Symmetric Initial Population Distribution ($N=10$, $L=6$ ) }
  \label{fig:4}
\end{figure}

\subsection{With Memory VS Without Memory}
 
 The first big test to the model is determining whether it works better than pure randomness. To do this 100 tries with the above initial conditions were made, and the days a population lasted were registered in a variable named $lifepop$. A try will last at maximum 50 days, since after this value is reached it's reasonable to conclude that the population will last 'forever'.
 
\begin{figure}[H] 
  \fbox{
    \centering
    \includegraphics[width=0.9\linewidth]{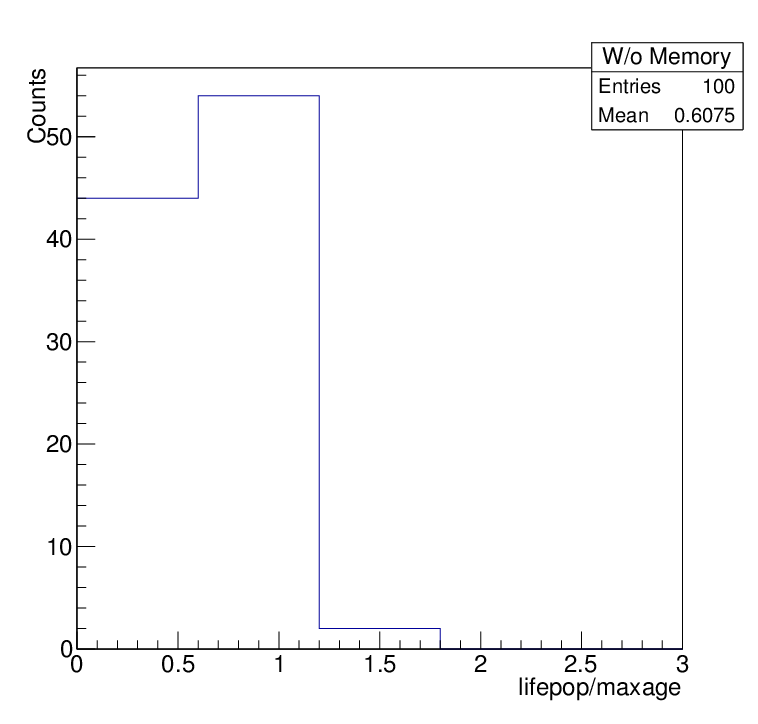} 
    }
    \label{fig:5a} 

  \fbox{ 
    \centering
    \includegraphics[width=0.9\linewidth]{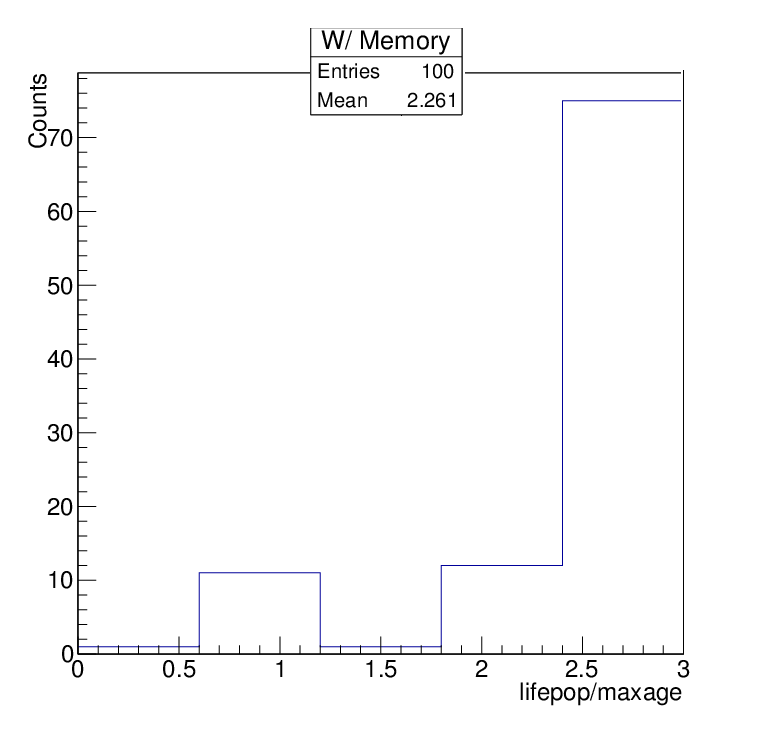} 
    }
    \label{fig:5b} 
  \caption{(Top) Tries Without Memory ; (Bottom) Tries with Memory}
  \label{fig:5} 
\end{figure}

Observing FIG. \ref{fig:5} we can conclude that the model works. As for without memory, most populations will last the maximum days a single \textit{cell} can live. Most of the times this corresponds to a single \textit{cell} that randomly ate every time it needed, while all the other died. As for the with memory case, most of the tries will last 'forever' while some unlucky scenarios are still possible (no food source is discovered in time in the beginning). Moreover, we can already detected some interesting tries that last long but not 'forever'. This will be related to why we named "Adventurous" evolutions, and putting it in simple terms, are adventures that went bad (more on that on the following section).

\subsection{Interesting Evolutions of the System}

A trivial evolution of the system is when the population discovers both food sites really early, reaching food sites saturation and their equilibrium number after few days. This evolution is named 'Lucky Evolution'.
The other presented evolution  was named 'Adventurous' because only one food site is discovered at the beginning, the population develops around it until it is saturated and will need to take risks in order to evolve - move away from the only food site it knows. \\

\subsubsection{Lucky Evolution}

\begin{figure}[H] 
\centering
 \fbox{
    \centering
    \includegraphics[width=0.8\linewidth]{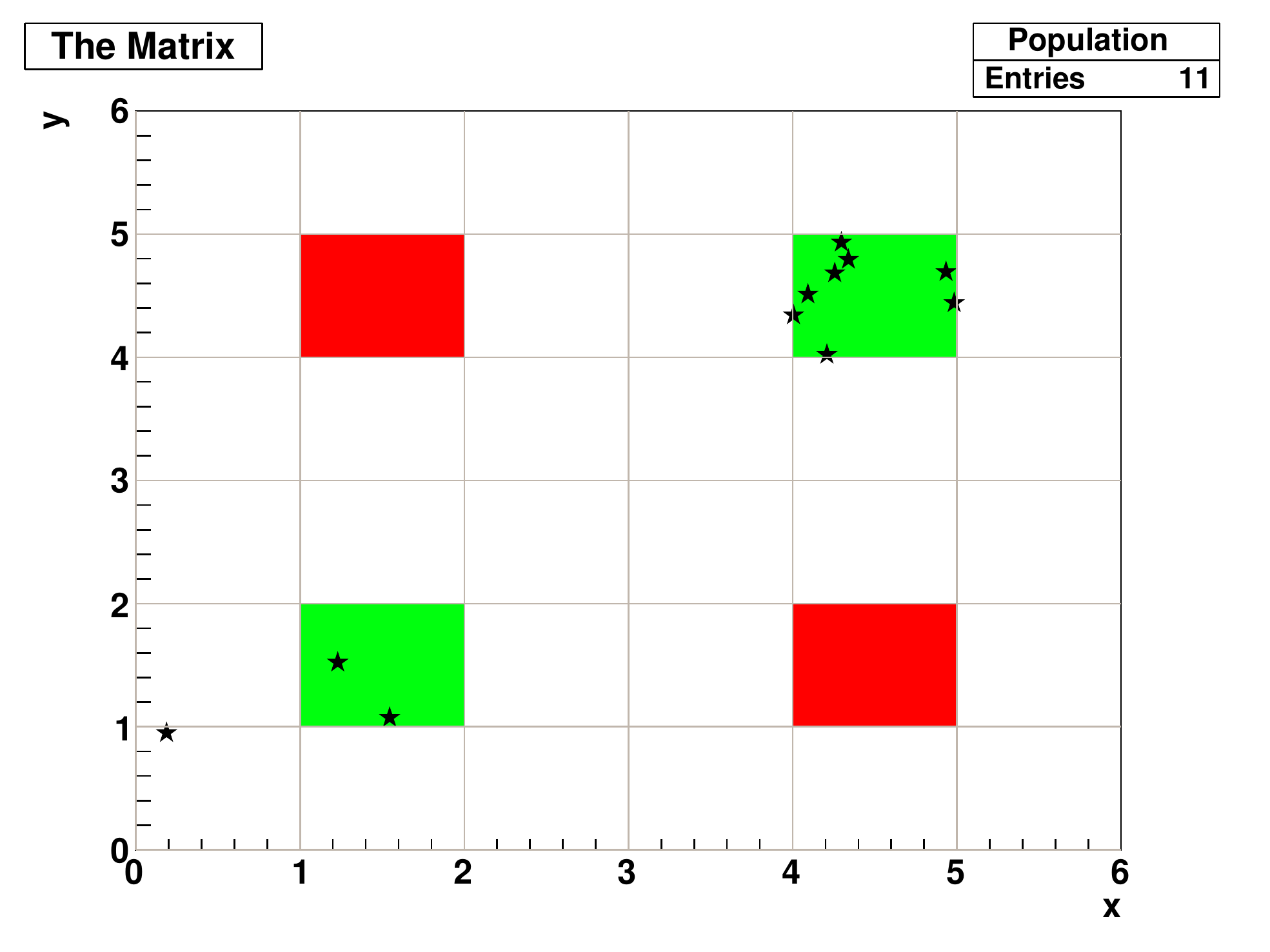} }
    \caption{Day 5} 
    \label{fig:6a} 
    \vspace{4ex}

 \fbox{
    \centering
    \includegraphics[width=0.8\linewidth]{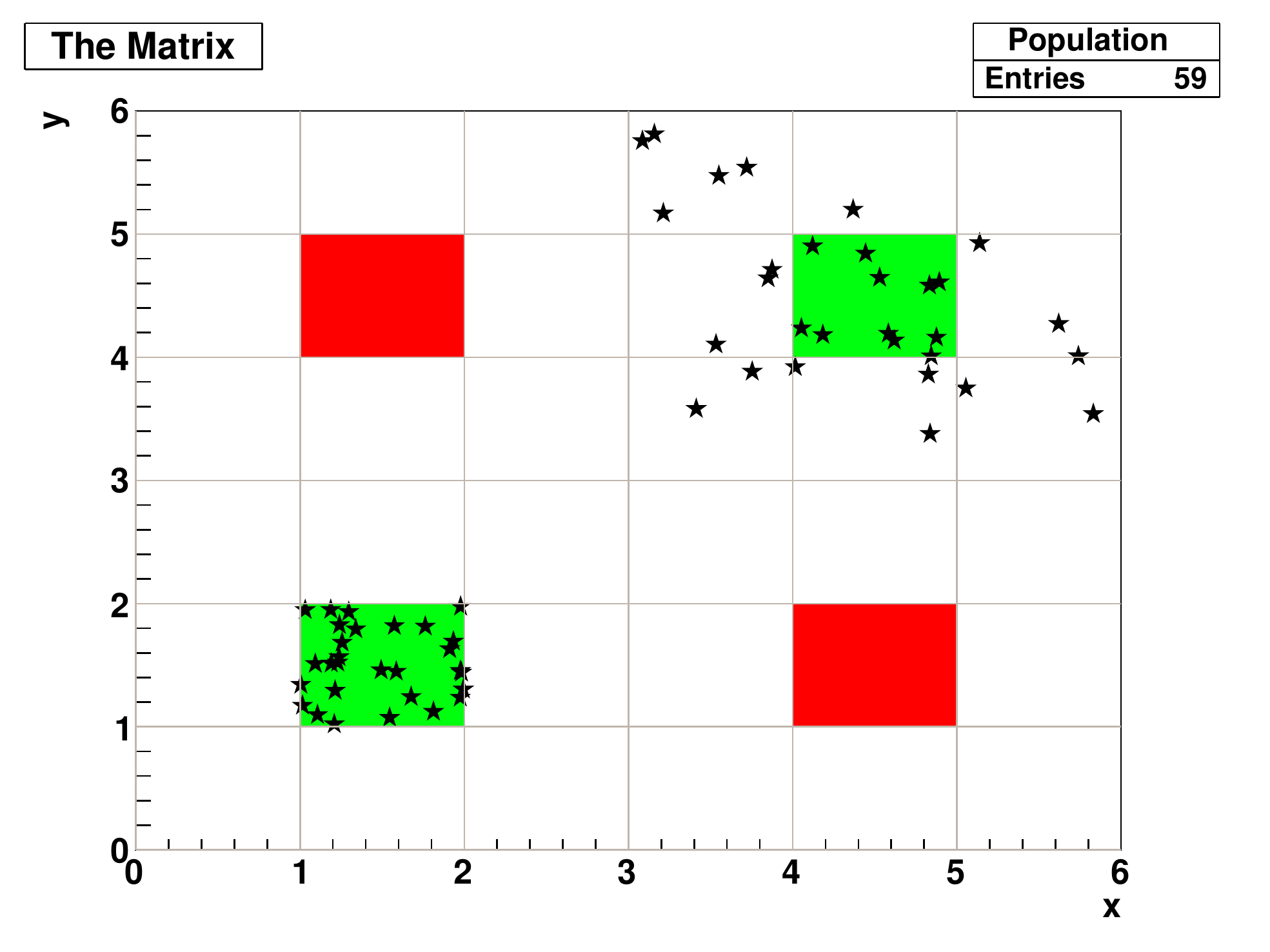} 
    }
    \caption{Day 14} 
    \label{fig:6b} 
    \vspace{4ex}

  \label{fig:6} 
\end{figure}

 An expected development around food sites is achieved.

\begin{figure}[H] 
\centering

 \fbox{
    \centering
    \includegraphics[width=0.8\linewidth]{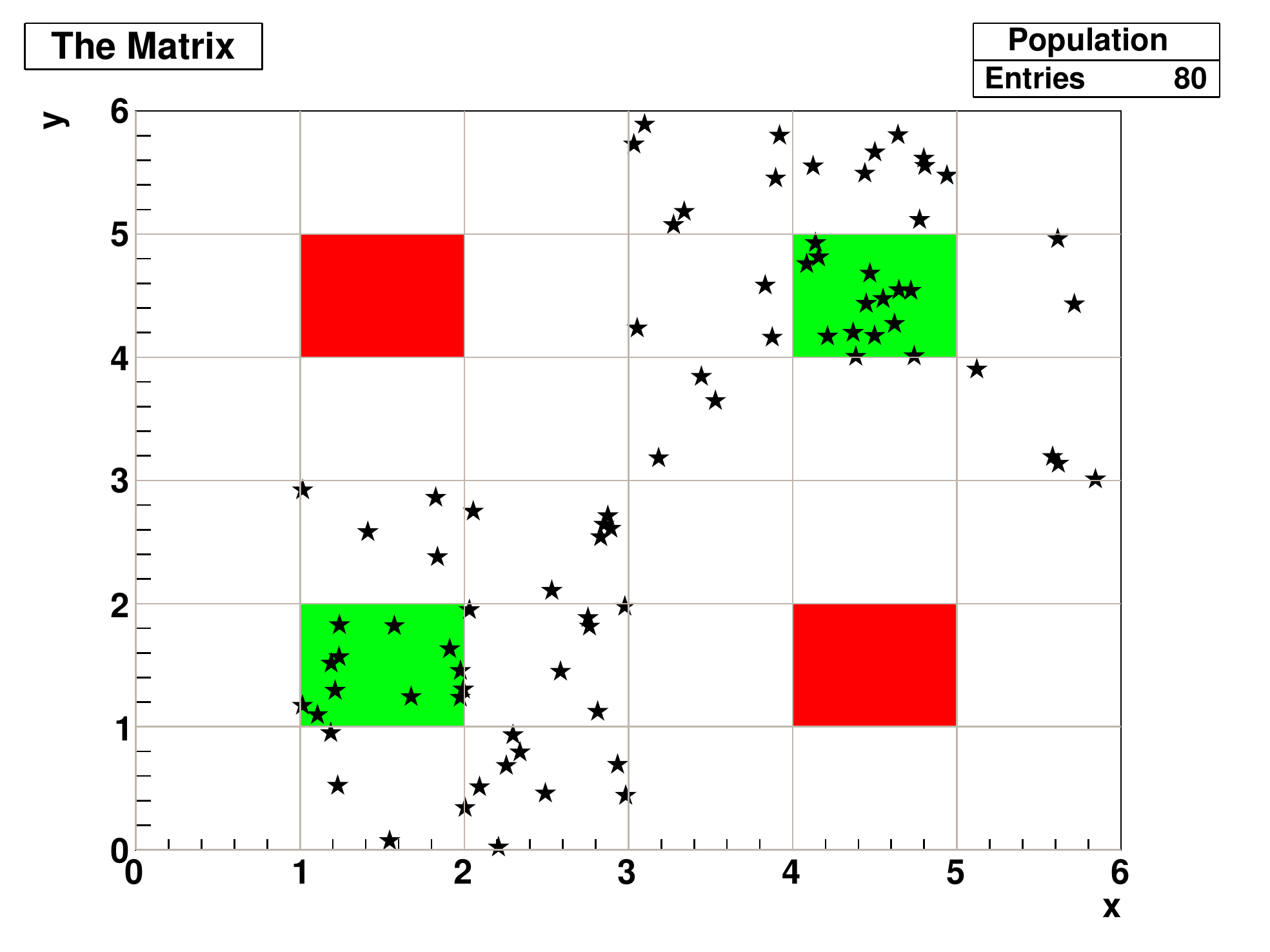} 
    }
    \caption{Day 58} 
    \label{fig7:a} 
    \vspace{4ex}

  \fbox{
    \centering
    \includegraphics[width=0.8\linewidth]{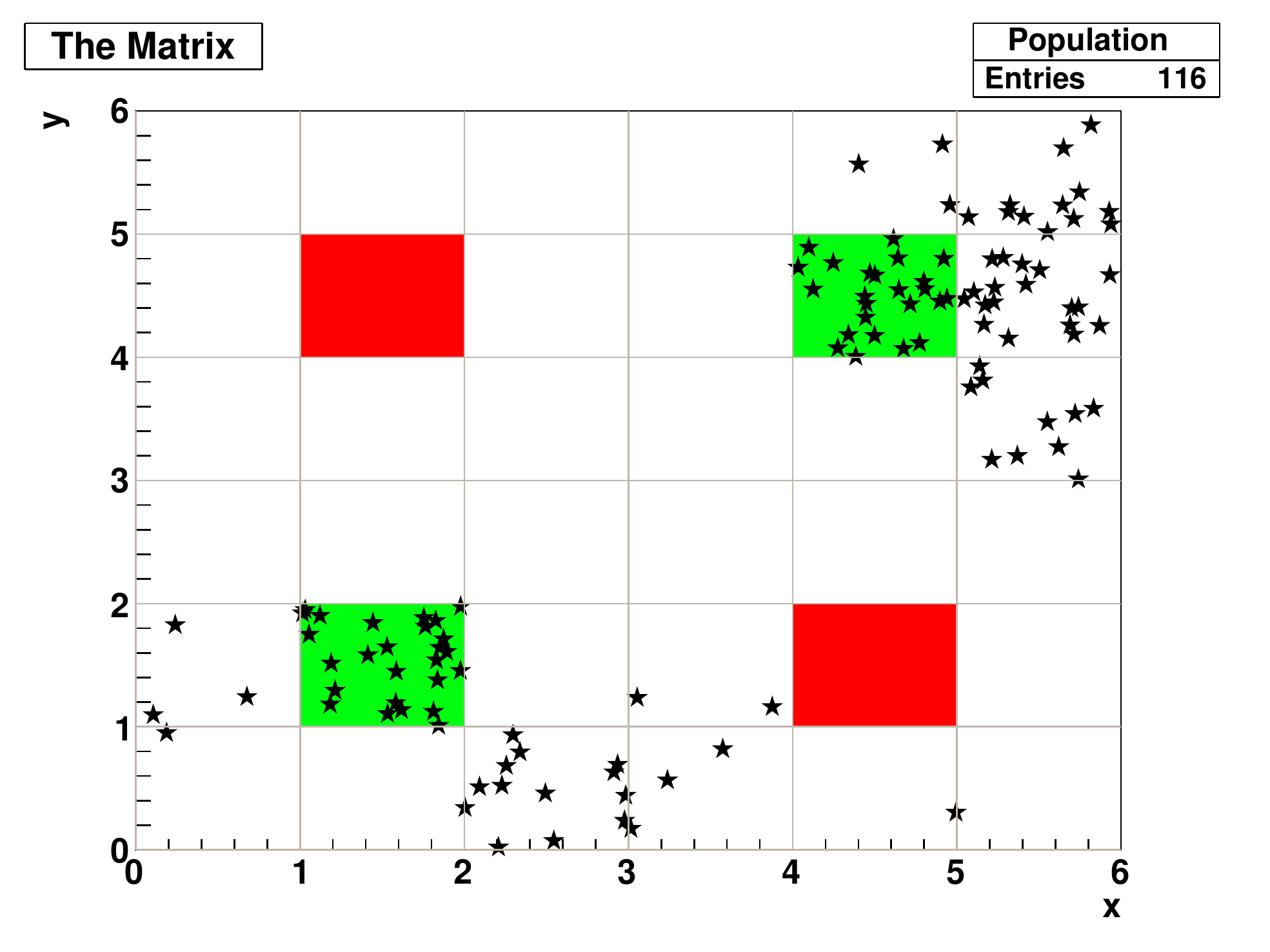} }
    \caption{Day 70} 
    \label{fig7:c}

  \label{fig:7} 
\end{figure}

An interesting (minimal) path between food sites is created on Day 58. We can say that the food sites exchange cells using it.

\begin{figure}[H]
\centering
  \includegraphics[scale=0.35]{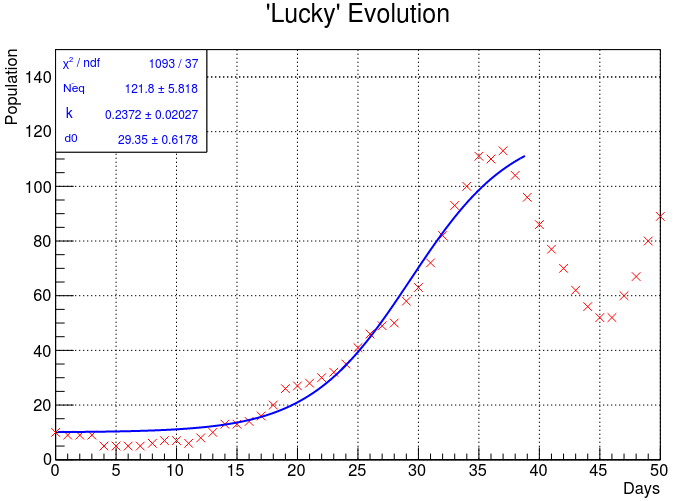}
\caption{Fit to equation  (\ref{eq:1}); \ $N_i=10$ \  }
  \label{fig:8}
\end{figure}

The reaching of the equilibrium can clearly be seen. After it is achieved the population starts looking for new food sites and this will cause the population number to decrease. After a while it starts increasing again, resulting in the observed oscillation.

\subsubsection{Adventurous Evolution}

\begin{figure}[H] 
 \fbox{
    \centering
    \includegraphics[width=0.8\linewidth]{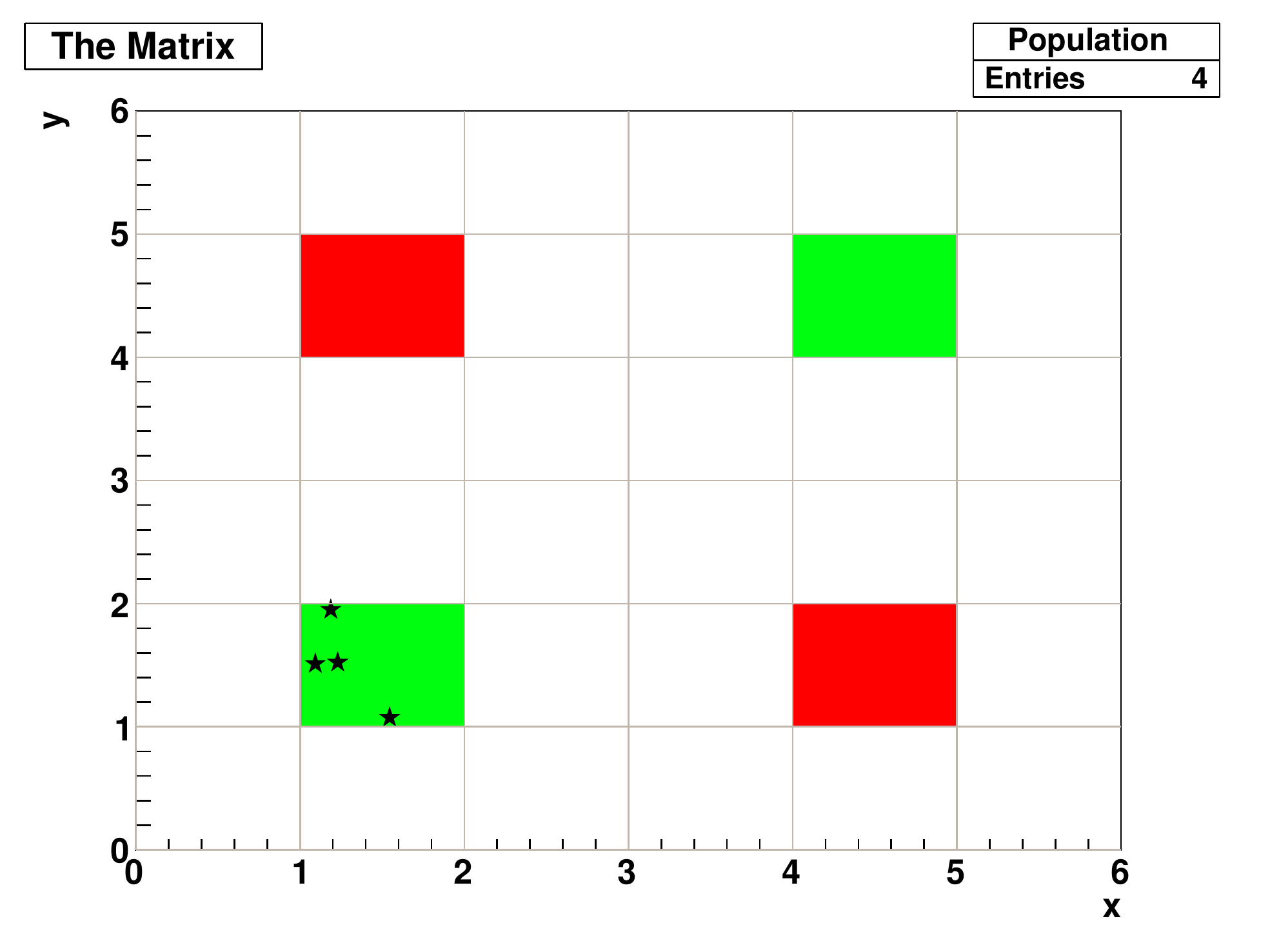} }
    \caption{Day 4} 
    \label{fig:6a} 
    \vspace{4ex}

 \fbox{
    \centering
    \includegraphics[width=0.8\linewidth]{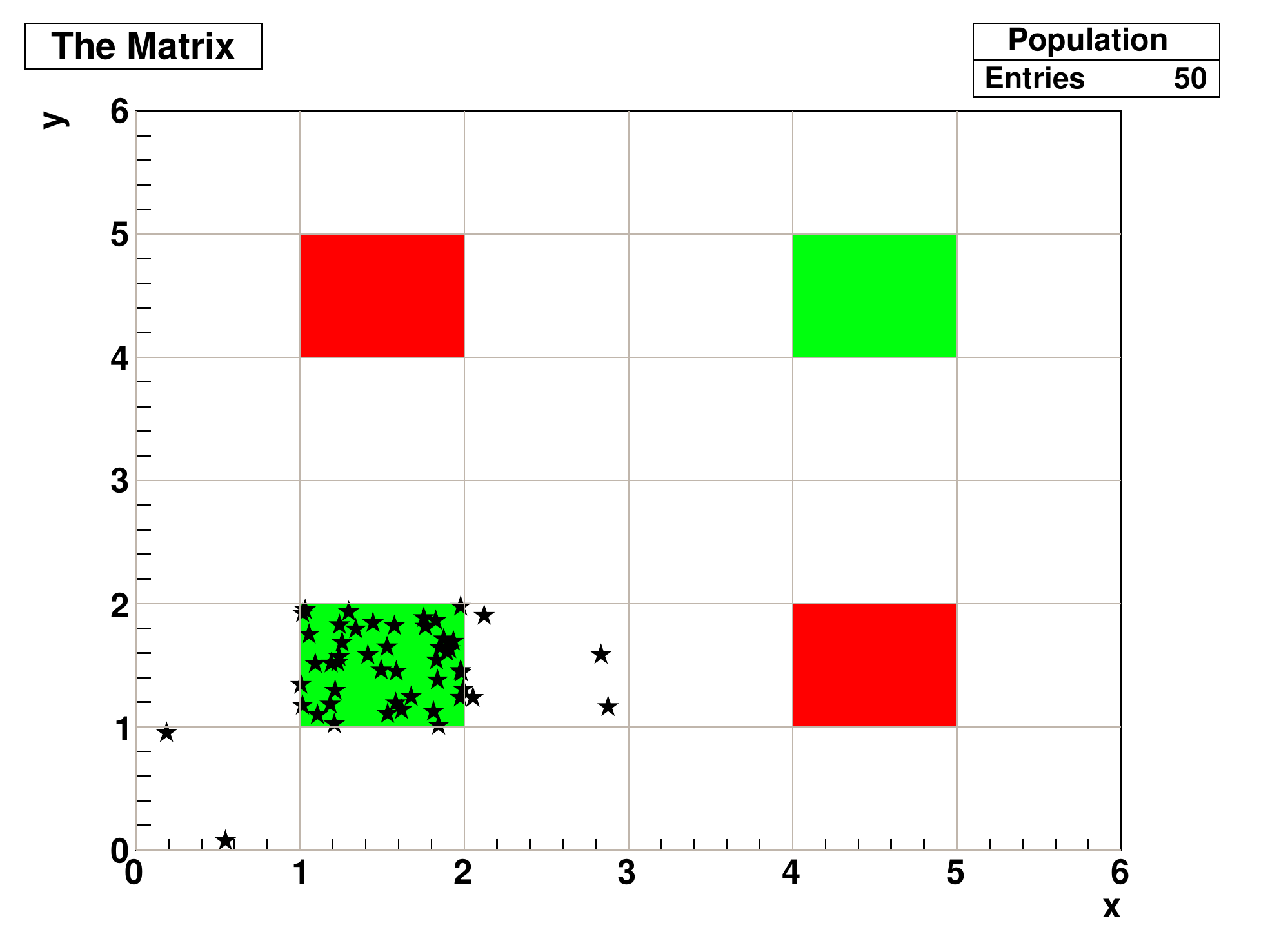} }
    \caption{Day 18} 
    \label{fig:6b} 
    \vspace{4ex}

 \fbox{
    \centering
    \includegraphics[width=0.8\linewidth]{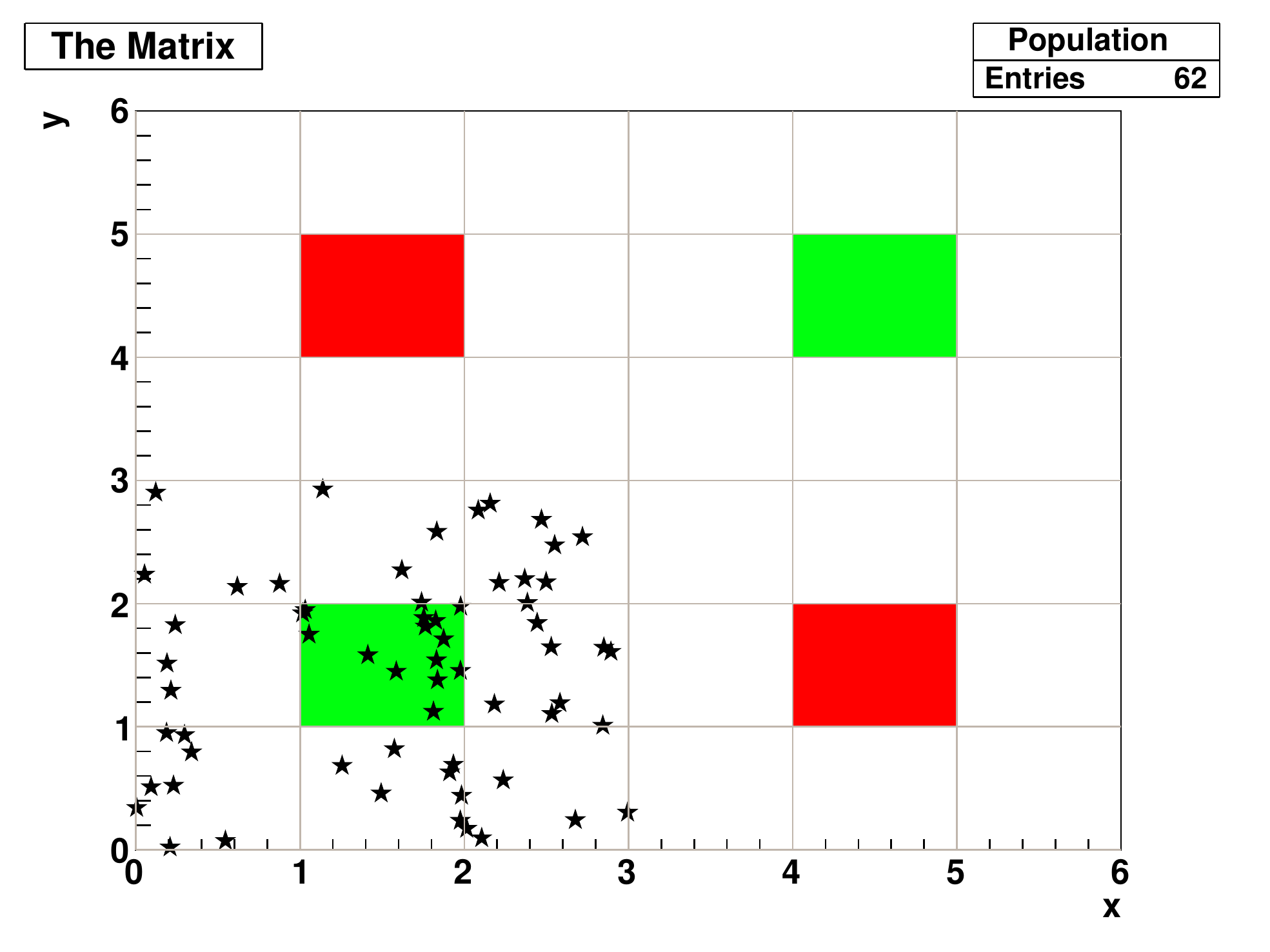} }
    \caption{Day 23} 
    \label{fig:6c}

  \label{fig:6} 
\end{figure}

 Then, a journey starts to find the other food site, where many \textit{cells} die, but their effort is not pointless since all is registered in the memory of the newer generations, that will eventually reach the new food site.

\begin{figure}[H] 
 \fbox{
    \centering
    \includegraphics[width=0.8\linewidth]{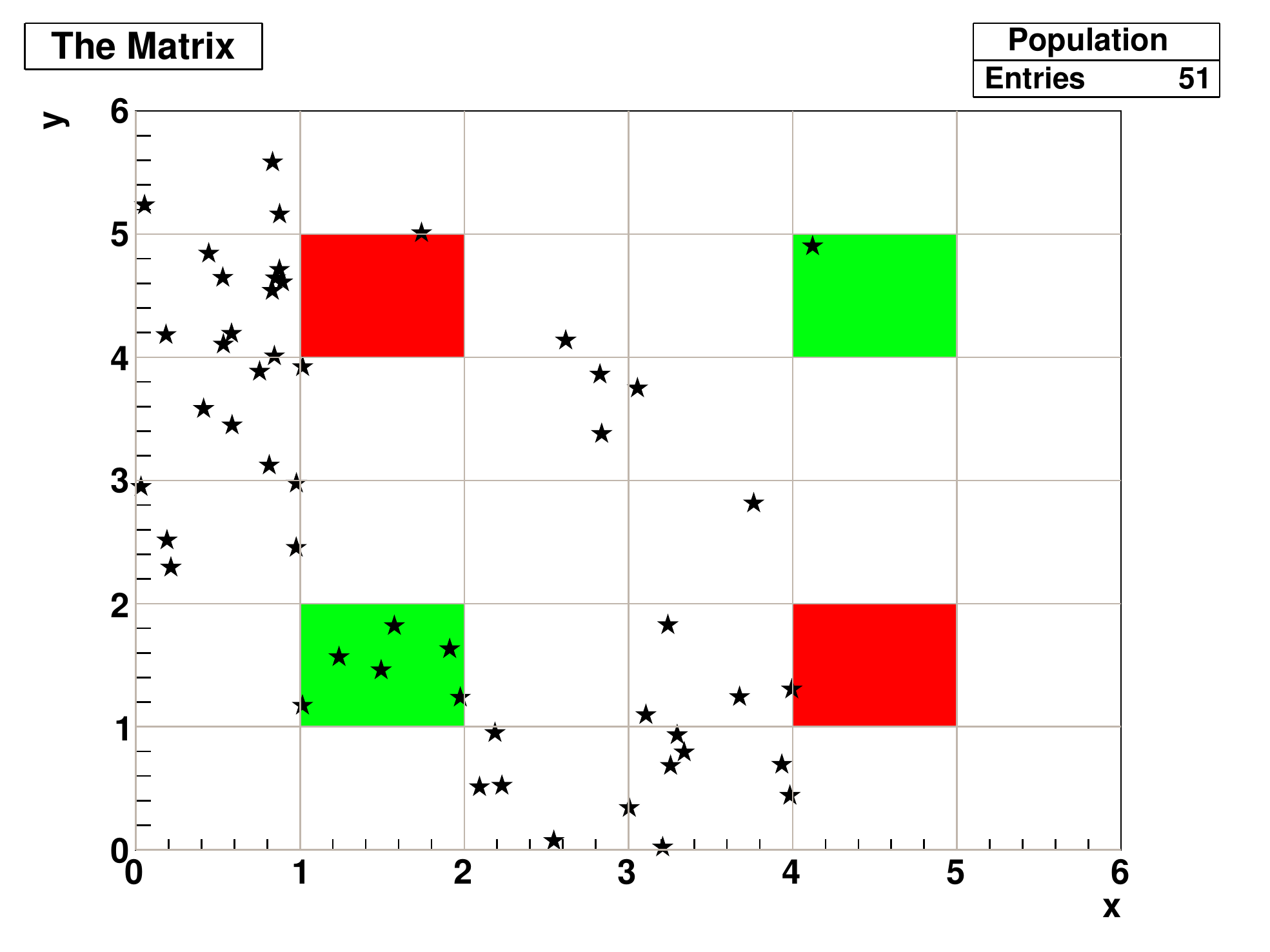} }
    \caption{Day 45} 
    \label{fig7:a} 
    \vspace{4ex}

 \fbox{
    \centering
    \includegraphics[width=0.8\linewidth]{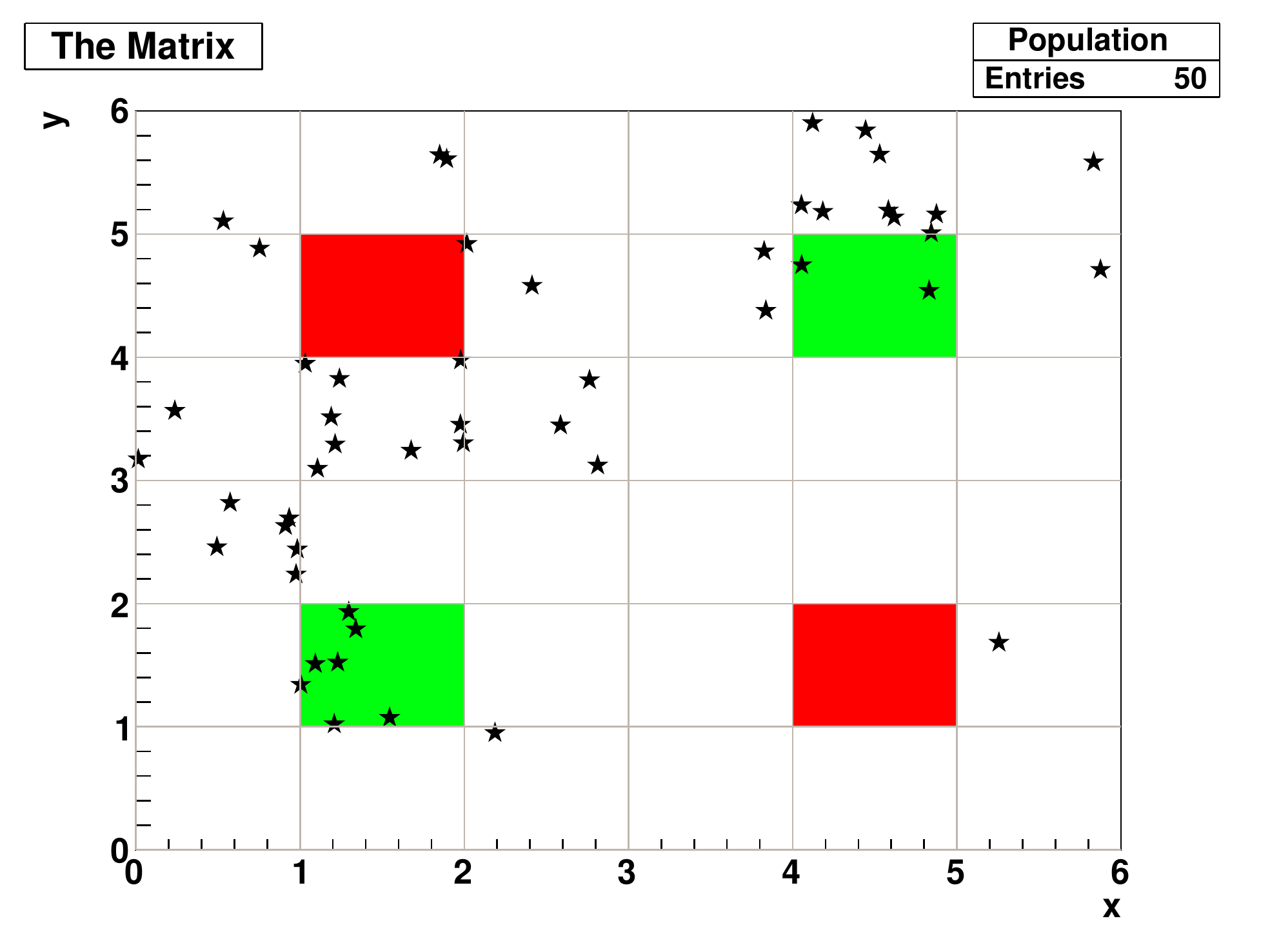} }
    \caption{Day 51} 
    \label{fig7:b} 
    \vspace{4ex}

 \fbox{
    \centering
    \includegraphics[width=0.8\linewidth]{a_70.pdf} 
    }
    \caption{Day 70} 
    \label{fig7:c} 
 
  \label{fig:7} 
\end{figure}
From now they will develop until both food sites are overflowed, and then the search for a new one restarts.

 \begin{figure}[H]
\centering
  \includegraphics[scale=0.35]{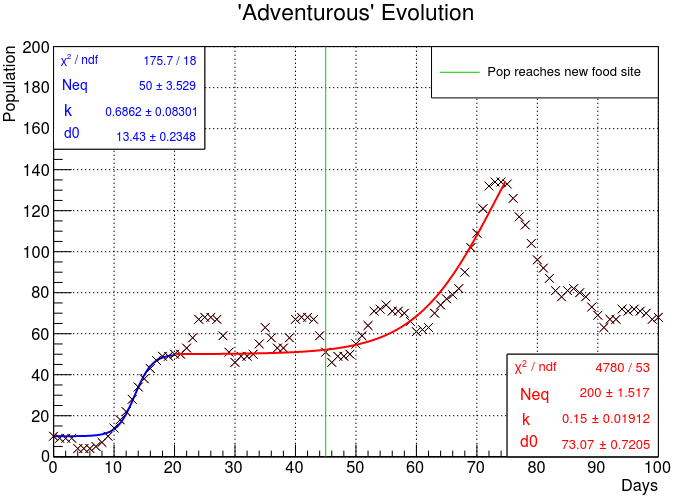}
\caption{Fit to equation  (\ref{eq:1}); \ $N_i=10$ \ for blue and   \ $N_i=50$ for red }
  \label{fig:8}
\end{figure}

This fit is highly illustrative of what was said. The reaching of the first equilibrium can clearly be seen, and taking the average value of the following oscillation (consequence of the exploration of unknown positions) we can see that the evolution towards the second equilibrium is of the same form as the first one.

\section{Conclusions}

The model used fulfilled its purpose of generating interesting collective behaviour having only three available tools: initial randomness, memory of mistakes and good moves and a tuning of the initial conditions.\\
The breakthrough that distinguished the memoryless case from the intelligent one  was the correct definition of probability distribution and of the quality function and gifting generation 0 with a live memory that emulates their only way of sharing information - communication. This effect can be neglected as soon as generation 1 is born, since the amount of data created by a previous generation is of much bigger magnitude than the data created daily.\\
The programming language used was C++ and the graphical objects were made using ROOT.\\
Natural Selection has 3 main ingredients: Heredity, Selection and Variation. Only the first two were used in this algorithm. An obvious improvement would be introducing mutations - randomly generated information added to the memory of some generations that can contribute in a good or bad way to the development.\\
Furthermore, an interesting test would be introducing a concurrent population with a different genetic pool (a different $mem_{tot}$ and observe if both populations tend to merge and add their memories.\\
This model can be compared to the behaviour of unicellular organisms but also of humans around major cities, correctly predicting their migration when deaths start to increase abruptly (war zones and/or zones with a food shortage).
\vspace{-6mm}
\acknowledgements

The author would like to thank Professor Fernando Barao for pushing his C language boundaries in the Computational Physics course, Professor Filipe R. Joaquim for valuable advices on the structure of the article and Professor Jean-Sebastien Caux from the University of Amsterdam for introducing the author to the random walk process.

\end{document}